\documentclass[journal,twocolumn]{IEEEtran}

\usepackage{palatino}
\usepackage{mathtools}
\usepackage{url}
\usepackage{amsmath}
\usepackage{amssymb}
\usepackage{wrapfig}
\usepackage{framed}
\usepackage{mdframed}
\usepackage{verbatim}
 \usepackage{amsthm}
 \usepackage{cite}
\usepackage{subfigure}
\usepackage{palatino}
\usepackage{mathtools}
\usepackage{url}
\usepackage{amsmath}
\usepackage{amssymb}
\usepackage{wrapfig}
\usepackage{framed}
\usepackage{mdframed}
\usepackage{bbm}
\usepackage{verbatim}\usepackage{subfigure}
\usepackage[utf8]{inputenc}
\usepackage{amssymb}
\usepackage{amsthm}
\usepackage{dsfont}
\usepackage[normalem]{ulem}

\usepackage{bbm}
\usepackage{subfigure}
\usepackage{times}
\usepackage{multirow}
\usepackage[final]{graphicx}
\usepackage{upgreek}
\usepackage{latexsym,amssymb,mathrsfs}
\usepackage{comment}
\usepackage{url}

\usepackage{algorithmicx}
\usepackage[ruled]{algorithm}
\usepackage{algpseudocode}
\usepackage{algpascal}
\usepackage{algc}
\usepackage{xcolor}
\usepackage{cite}

 \usepackage{cite}

\newtheorem{defi}{Definition}

\title{Local Privacy-preserving Mechanisms and Applications in Machine Learning}
\author{\IEEEauthorblockN{Likun~Qin,  Tianshuo Qiu \\
}
\IEEEauthorblockA{Department of Electrical and Computer Engineering}\\
\IEEEauthorblockA{Shandong University, Jinan, China\\
}
}
\begin{document}
\maketitle

\begin{abstract}

The emergence and evolution of Local Differential Privacy (LDP) and its various adaptations play a pivotal role in tackling privacy issues related to the vast amounts of data generated by intelligent devices, which are crucial for data-informed decision-making in the realm of crowdsensing. Utilizing these extensive datasets can provide critical insights but also introduces substantial privacy concerns for the individuals involved. LDP, noted for its decentralized framework, excels in providing strong privacy protection for individual users during the stages of data collection and processing. The core principle of LDP lies in its technique of altering each user's data locally at the client end before it is sent to the server, thus preventing privacy violations at both stages.  There are many LDP variances in the privacy research community aimed to improve the utility-privacy tradeoff. On the other hand, one of the major applications of the privacy-preserving mechanisms is machine learning. In this paper, we firstly delves into a comprehensive analysis of LDP and its variances, focusing on their various models, the diverse range of its adaptations, and the underlying structure of privacy mechanisms; then we discuss the state-of-art privacy mechanisms applications in machine learning.
\end{abstract}

\section{Introduction}

Collecting and analyzing data introduces significant privacy concerns because it often includes sensitive user information. With the advent of sophisticated data fusion and analysis methods, user data becomes even more susceptible to breaches and exposure in this era of big data. For instance, by studying appliance usage, adversaries can deduce daily routines or behaviors of individuals, like when they are home or their specific activities such as watching TV or cooking. It's crucial to prioritize the protection of personal data when gathering information from diverse devices. Currently, the European Union (EU) has released the GDPR\cite{GDPR2016a}, which oversees EU data protection laws for its citizens and outlines the specifics related to the handling of personal data. Similarly, the U.S. National Institute of Standards and Technology (NIST) is in the process of crafting privacy frameworks. These frameworks aim to more effectively recognize, evaluate, and address privacy risks, enabling individuals to embrace innovative technologies with increased trust and confidence\cite{Facebook,free_lunch}.

From a privacy-protection standpoint, differential privacy (DP) has been introduced over a decade ago \cite{Dwork2006, Dwork20061}. Recognized as a robust framework for safeguarding privacy, it's often termed as global DP or centralized DP. DP's strength lies in its mathematical rigor; it operates independent of an adversary's background knowledge and assures potent privacy protection for users. It has found applications across various domains\cite{DP_mec}. However, DP assumes the presence of a trustworthy server, which can be a challenge since many online platforms or crowdsourcing systems might have untrustworthy servers keen on user data statistics\cite{primoff2017equifax, lu2019assessing}.

Emerging from the concept of DP, local differential privacy (LDP) was introduced \cite{Dwork2008}. LDP stands as a decentralized version of DP, offering individualized privacy assurances and making no assumptions about third-party server trustworthiness. LDP has become a focal point in privacy research due to its theoretical significance and practical implications\cite{Rappor}. Numerous corporations, including Apple's iOS\cite{apple_ldp}, Google Chrome, and the Windows operating system, have integrated LDP-driven algorithms into their systems. Owing to its robust capabilities, LDP has become a preferred choice to address individual privacy concerns during various statistical and analytical operations. This includes tasks like frequency and mean value estimation\cite{Tianhao}, the identification of heavy hitters\cite{heavyhitter}, k-way marginal release, empirical risk minimization (ERM), federated learning, and deep learning.

While LDP is powerful, it's not without its challenges, notably in striking an optimal balance between utility and privacy\cite{context}. To address this, there are two primary approaches. Firstly, by devising improved mechanisms - leading to the introduction of numerous LDP-based protocols and sophisticated mechanisms in academic circles. Secondly, by revisiting the definition of LDP itself, with researchers suggesting more flexible privacy concepts to better cater to the utility-privacy balance required for real-world applications. Given the growing significance of LDP, a thorough survey of the topic is both timely and essential. While there exists some literature reviewing LDP, the focus has often been narrow. They either focus on specific applications or certain types of mechanisms.  

In this paper, we delve deep into the world of LDP and its various offshoots, meticulously studying their recent advancements and associated mechanisms. We embark on a thorough exploration of the foundational principles that drive LDP and the evolutionary trajectories of its multiple variants. We aim to identify the cutting-edge developments, shedding light on the innovations that have shaped these privacy tools and the challenges they aim to address in our contemporary digital landscape. Furthermore, we analyze the specific mechanisms that support and enhance the capabilities of LDP, understanding their technical intricacies and the real-world applications they cater to. Through this comprehensive study, we aspire to provide readers with a panoramic view of the current state of LDP research, setting the stage for future inquiries and innovations in this critical domain.

Recognizing the complexity of privacy issues in the digital age, we delve into how LDP mechanisms have been tailored and adapted to meet diverse needs and environments. We explore the balance between privacy preservation and the practical utility of data, a central challenge in the deployment of LDP.

Our survey extends to the innovative methodologies that have been developed within the LDP framework. These include refined algorithms and protocols that enhance privacy protection while maintaining data usefulness. We investigate how these advancements have been implemented in various real-world scenarios, from online platforms to smart device ecosystems, and their effectiveness in protecting user privacy.

Moreover, the paper examines the evolution of LDP in response to emerging technologies and changing data landscapes. We discuss how the concept of LDP has been broadened or altered to accommodate new types of data collection and analysis methods, including the growing field of federated learning and the increasing reliance on deep learning algorithms. In addressing these aspects, we also highlight the legal and ethical considerations surrounding LDP. The paper considers how LDP aligns with global data protection regulations like the GDPR and how it is influencing policy-making and privacy standards worldwide.

Finally, we present the state-of-the-art applications of the privacy models in machine learning. Specifically, we focus on supervised and unsupervised machine learning, Empirical Risk Minimization, Reinforcement Learning, Deep learning, and Federated Learning.

\section{Local Differential Privacy, Properties and Mechanisms}

This section explores Local Differential Privacy (LDP) and its associated mechanisms. We begin with an understanding of LDP's definition.

Definition 1: $\varepsilon$-Local Differential Privacy ($\varepsilon$-LDP) -
A randomized mechanism $M$ adheres to $\varepsilon$-LDP if, for any two input values $v$ and $v'$ within $M$'s domain, and for any output $y$ in $Y$, the following holds:
\begin{equation}
P[M(v) = y] \leq e^{\varepsilon} \times P[M(v') = y],
\end{equation}
where $P[\cdot]$ signifies probability, and $\varepsilon$ represents the privacy budget. Smaller $\varepsilon$ values imply stronger privacy.

LDP's key characteristics are:

Composition -
If two mechanisms, $M_1$ and $M_2$, offer $\varepsilon_1$-LDP and $\varepsilon_2$-LDP respectively, their combined use results in $(\varepsilon_1 + \varepsilon_2)$-LDP.
\begin{equation}
M(v) = (M_1(v), M_2(v)) \Rightarrow M \text{ provides } (\varepsilon_1 + \varepsilon_2) \text{-LDP}
\end{equation}

Post-processing Immunity -
Functions applied to an $\varepsilon$-LDP mechanism's output maintain the $\varepsilon$-LDP property.
\begin{equation}
\text{If } M(v) \text{ offers } \varepsilon\text{-LDP, then } f(M(v)) \text{ also ensures } \varepsilon\text{-LDP}.
\end{equation}

Resistance to Auxiliary Information -
LDP remains effective even if adversaries possess additional information.

Utility-Privacy Balancing -
A lower $\varepsilon$ generally means increased privacy but can reduce the data's usefulness. Independence from Background Knowledge -
LDP's privacy assurances are unaffected by an adversary's prior knowledge. Subsequently, we examine LDP-compliant mechanisms:
Randomized Response Mechanism -
This technique, effective for binary data, involves a user responding truthfully or randomly with equal probability. Its probability mass function (pmf) ensures $\varepsilon$-LDP with $\varepsilon = \ln(2)$. Laplace Mechanism -
This approach adds Laplace-distribution noise to data, with the noise level determined by the sensitivity of the function and $\varepsilon$. Gaussian Mechanism -
Similar to the Laplace Mechanism, but uses Gaussian distribution noise. The noise amount is based on the desired $\varepsilon$ and function sensitivity. Exponential Mechanism -
Selects outputs based on a scoring function, with selection probability proportional to the exponential of their score. Perturbed Histogram Mechanism -
Rather than altering each item, this method modifies the entire histogram of data items, adding Laplace-distribution noise. The effectiveness of each mechanism is closely tied to the query's sensitivity, denoted as $\Delta f$. In LDP, high sensitivity can necessitate substantial noise, potentially reducing data utility.

Increasing input support size complicates maintaining desired privacy levels, as high sensitivity noise can obscure actual data, leading to potential misinterpretation or meaningless results. This highlights a significant tradeoff between data utility and privacy. Stronger privacy often comes at the cost of accuracy and utility, posing challenges for precision-requiring applications. While LDP offers robust theoretical privacy, its practical application demands a delicate balance between utility and privacy, spurring ongoing research for more effective mechanisms or modified privacy models.

\section{LDP variants and mechanisms}
In this section, we introduce LDP variants that aim to provide better utility-privacy tradeoff in different applications.
\subsection{Variants and Mechanisms of LDP}

\subsubsection{\( (\varepsilon, \delta) \)-LDP}
Drawing parallels with how \( (\varepsilon, \delta) \)-DP \cite{DBLP:journals/corr/abs-1810-02810} extends \( \varepsilon \)-DP, \( (\varepsilon, \delta) \)-LDP (sometimes termed as approximate LDP) serves as a more flexible counterpart to \( \varepsilon \)-LDP (or pure LDP). 

\begin{defi}[Approximate Local Differential Privacy]
A randomized process \( M \) complies with \( (\varepsilon, \delta) \)-LDP if, for all input pairs \( v \) and \( v' \) within \( M \)'s domain and any probable output \( y \in Y \), the following holds:
\[ P[M(v) = y] \leq e^{\varepsilon} \cdot P[M(v') = y] + \delta. \]
Here, \( \delta \) is customarily a small value.
\end{defi}

In essence, \( (\varepsilon, \delta) \)-LDP implies that \( M \) achieves \( \varepsilon \)-LDP with a likelihood not less than \( 1-\delta \). If \( \delta = 0 \), \( (\varepsilon, \delta) \)-LDP converges to \( \varepsilon \)-LDP.

\subsubsection{BLENDER Model}
BLENDER \cite{203630}, a fusion of global DP and LDP, optimizes data utility while retaining privacy. It classifies users based on their trust in the aggregator into two categories: the opt-in group and clients. BLENDER enhances utility by balancing data from both. Its privacy measure mirrors that of \( (\varepsilon, \delta) \)-DP \cite{DBLP:conf/eurocrypt/DworkKMMN06}.

\subsubsection{Geo-indistinguishability}
Originally tailored for location privacy with global DP, Geo-indistinguishability \cite{Geo} uses the data's geographical distance. Alvim et al. \cite{DBLP:journals/corr/abs-1805-01456} argued for metric-based LDP's advantages in specific contexts.

\begin{defi}[Geo-indistinguishability]
A randomized function \( M \) adheres to Geo-indistinguishability if, for any input pairs \( v \) and \( v' \) and any output \( y \in Y \), the subsequent relation is met:
\[ P[M(v) = y] \leq e^{\varepsilon \cdot d(v, v')} \cdot P[M(v') = y], \]
where \( d(., .) \) designates a distance metric.
\end{defi}

This model adjusts privacy depending on data distance, augmenting utility for datasets like location or smart meter consumption that are sensitive to distance.

\subsubsection{Local Information Privacy}
Local Information Privacy (LIP) was originally proposed in \cite{Jian1805:Context} as a prior-aware version of LDP, and then, in \cite{jiang2019local}, Jiang et al relax the prior-aware assumption to partial prior-aware (Bounded Prior in their version). The definition of LIP is shown as follows: 

\begin{defi}

$(\epsilon,\delta)$-Local Information Privacy\cite{LIP1}
A mechanism ${M}$  satisfies $(\epsilon,\delta)$-LIP, if $\forall{x\in{\mathcal{X}}}$, $y\in{\textit{Range}(\mathcal{M})}$:
%\begin{equation}\label{def:LIP}
%    e^{-\epsilon}P(Y\in{\mathcal{S}_y})\le{P(Y\in\mathcal{S}_y|X\in\mathcal{S}_x)}\le{e^{\epsilon}}P(Y\in{\mathcal{S}_y})+\delta.
%\end{equation}
\begin{equation}\label{cons0}
\begin{aligned}
    & P(Y=y) \geq e^{-\epsilon}P(Y=y|X=x)-\delta, \\
    &P(Y=y)\le{e^{\epsilon}P(Y=y|X=x)}+\delta.
\end{aligned}
\end{equation}
    
\end{defi}

% \subsubsection{Sequential Information Privacy}

% Sequential Information Privacy (SIP), built upon LIP, measures the privacy leakage for a data sequence, or time series data. SIP naturally decomposes using chain rule-similar techniques and is comparable to that of LDP.

% \begin{defi}

% $[(\epsilon)$-Sequence Information Privacy]\cite{jiang2023online}
% A mechanism $\mathcal{M}$  satisfies $(\epsilon)$-SIP for some $\epsilon\in{\mathds{R}^+}$, if $\forall{X_1^T\in{\mathcal{X}}}$, $Y_1^T\in{\textit{Range}(\mathcal{M})}$:
% \begin{equation}\label{cons0}
% \begin{aligned}
%     e^{-\epsilon}\le \frac{P[M(x_1^T)=y_1^T]}{P[X_1^T=x_1^T]}\le{e^{\epsilon}}
% \end{aligned}
% \end{equation}
    
% \end{defi}

 \noindent The operational meaning of LIP is, the output $Y$ provides limited additional information about any possible input $X$, and the amount of the additional information is measured by the privacy budget $\epsilon$ and failure probability $\delta$. 

 In \cite{LIP2}, multiple LIP mechanisms were proposed and testified, showing that even though $\epsilon$-LIP is stronger than $2\epsilon$-LDP in terms of privacy protection. The mechanisms achieve more than 2 times of utility gain.
 
\subsubsection{CLDP}
Recognizing LDP's diminished utility with fewer users, Gursoy et al. \cite{DBLP:journals/corr/abs-1905-06361} introduced the metric-based model of condensed local differential privacy (CLDP).

\begin{defi}[\( \alpha \)-CLDP]
For all input pairs \( v \) and \( v' \) in \( M \)'s domain and any potential output \( y \in Y \), a randomized function \( M \) satisfies \( \alpha \)-CLDP if:
\[ P[M(v) = y] \leq e^{\alpha \cdot d(v, v')} \cdot P[M(v') = y], \]
where \( \alpha > 0 \).
\end{defi}

In CLDP, a decline in \( \alpha \) compensates for a growth in distance \( d \). Gursoy et al. employed an Exponential Mechanism variant to devise protocols, particularly benefitting scenarios with limited users.

% \subsubsection{Mutual Information Privacy}

\subsubsection{PLDP}
PLDP \cite{8368271} offers user-specific privacy levels. Here, users can modify their privacy settings, denoted by \( \varepsilon \).

\begin{defi}[\( \varepsilon \)-PLDP]
For a user \( U \), and all input pairs \( v \) and \( v' \) in \( M \)'s domain and any potential output \( y \in Y \), a randomized function \( M \) meets \( \varepsilon_U \)-PLDP if:
\[ P[M(v) = y] \leq e^{\varepsilon_U} \cdot P[M(v') = y]. \]
\end{defi}

Approaches like the personalized count estimation protocol and advanced combination strategy cater to users with varying privacy inclinations.

\subsubsection{Utility-optimized LDP (ULDP)}

Traditional LDP assumes all data points have uniform sensitivity, often causing excessive noise addition. Recognizing that not all personal data have equivalent sensitivity, the Utility-optimized LDP (ULDP) model was proposed. In this model, let \( KS \subseteq K \) be the sensitive dataset and \( KN = K \setminus KS \) be the non-sensitive dataset. Let \( Y_P \subseteq Y \) be the protected output set and \( Y_I = Y \setminus Y_P \) be the invertible output set. The formal definition of ULDP is:

\begin{defi}
Given \( KS \subseteq K \), \( Y_P \subseteq Y \), a mechanism \( M \) adheres to \( (KS, Y_P, \epsilon) \)-ULDP if:
\begin{itemize}
    \item For every \( y \in Y_I \), there is a \( v \in KN \) with \( P[M(v) = y] > 0 \) and \( P[M(v') = y] = 0 \) for any \( v' \neq v \).
    \item For all \( v, v' \in K \) and \( y \in Y_P \), \( P[M(v) = y] \leq e^{\epsilon} \cdot P[M(v') = y] \).
\end{itemize}
\end{defi}

In simpler terms, \( (KS, Y_P, \epsilon) \)-ULDP ensures that sensitive inputs are mapped only to the protected output set.

\subsubsection{Input-Discriminative LDP (ID-LDP)}

While ULDP classifies data as either sensitive or non-sensitive, the ID-LDP model offers a more nuanced approach by acknowledging varying sensitivity levels among data. It is defined as:

\begin{defi}
Given a set of privacy budgets \( E = \{\epsilon_v\}_{v\in K} \), a mechanism \( M \) adheres to \( E \)-ID-LDP if for all input pairs \( v \) and \( v' \), and any output \( y \in Y \):
\[ P[M(v) = y] \leq e^{r(\epsilon_v, \epsilon_{v'})} \cdot P[M(v') = y] \]
where \( r(\cdot, \cdot) \) is a function of two privacy budgets.
\end{defi}

The study in \cite{DBLP:journals/corr/abs-1807-11317} primarily utilizes the minimum function between \( \epsilon_v \) and \( \epsilon_{v'} \) and introduces the MinID-LDP as a specialized case.

\subsubsection{Parameter Blending Privacy (PBP)}

PBP was proposed as a more flexible LDP variant \cite{10.1145/3320269.3405441}. In PBP, let \( \Theta \) represent the domain of privacy parameters. Given a privacy budget \( \theta \in \Theta \), let \( P(\theta) \) denote the frequency with which \( \theta \) is selected. PBP is defined as:

\begin{defi}
A mechanism \( M \) adheres to \( r \)-PBP if, for all \( \theta \in \Theta, v, v' \in K, y \in Y \), there exists a \( \theta' \in \Theta \) such that:
\[ P(\theta)P[M(v; \theta) = y] \leq e^{r(\theta)} \cdot P(\theta')P[M(v'; \theta') = y] \]
\end{defi}

\subsubsection{Perfect Privacy}

Another variant yet widely investigated privacy notion is perfect privacy. Perfect privacy can be understood as a special case of DP(LDP) when $\epsilon = 0$, or simply, the input is independent of the output.
\begin{defi}
    A mechanism \( M \) achieves perfect privacy for the input if, for all \( x \in \mathcal{X} \):
\begin{equation*}
    P(X|Y)=P(X).
\end{equation*}
\end{defi}
From the information perspective, perfect privacy can also be understood as the mutual information between the input and the output is zero. Since perfect privacy means the input and the output is independent, it provides no utility in general. However, recently there are several adaptions of perfect privacy mechanisms that protecting a latent variable of the input. Hence, when the latent variable is independent of the output, the input can still be correlated to it and therefore provides certain utility\cite{10.1145/1142351.1142375, 10159417}.

\section{Local Privacy Mechanisms for Machine learning}
Machine learning, a critical method in data analysis, finds application across various fields. However, it's susceptible to several types of attacks during the training process. These include membership inference attacks \cite{7958568}, model inversion attacks \cite{10.1145/2810103.2813677}, and memorizing model attacks \cite{10.1145/3133956.3134077}. For instance, adversaries could exploit these vulnerabilities to access sensitive user data, as demonstrated in cases where memorized information is extracted during training \cite{7958568}. A particularly concerning example by Fredrikson et al. \cite{10.1145/2810103.2813677} highlighted how facial recognition systems can be compromised under model inversion attacks, thereby exposing the frailties of a trained machine learning model.

Extensive research has been conducted on machine learning algorithms incorporating global Differential Privacy (DP) through private training \cite{10.1145/2976749.2978318, 8683991, 10.1145/3219819.3220076}. The advent of Local Differential Privacy (LDP) further led to investigations into machine learning algorithms under the LDP framework, aiming to enhance privacy protection in a distributed manner. The subsequent subsections provide a comprehensive overview of existing machine learning algorithms that utilize LDP. These algorithms are categorized and summarized from various perspectives, including supervised learning, unsupervised learning, empirical risk minimization, deep learning, reinforcement learning, and federated learning.

\subsection{Unsupervised Learning}
Clustering, a fundamental problem in data analysis, has been explored under centralized Differential Privacy (DP) \cite{nissim2016locating, su2016differentially}. In the realm of Local Differential Privacy (LDP), Nissim and Stemmer \cite{nissim2018clustering} studied 1-clustering by identifying a minimum enclosing ball. Additionally, Sun et al. \cite{sun2019distributed} delved into non-interactive clustering under LDP, extending the Bit Vector mechanism from \cite{karapiperis2017distance}. They enhanced the encoding process and introduced the $\mathrm{kCluster}$ algorithm for clustering in an anonymous space. Further, Li et al. \cite{li2017local} proposed a mechanism for local-clustering-based collaborative filtering, employing the kNN algorithm to categorize items while ensuring item-level privacy.

In local clustering, individuals randomize their data before sharing it with an untrusted data curator, aiming for stronger privacy protection. While local clustering may not be as accurate as its central counterpart, it offers enhanced user privacy and is more adaptable to varied privacy specifications, like personalized privacy parameters \cite{akter2017computing}. Xia et al. \cite{xia2020distributed} applied LDP to K-means clustering, perturbing user data directly. They introduced a budget allocation scheme to mitigate noise and enhance accuracy. Despite these advances, research on clustering under LDP is still in its nascent stages.

\subsection{Supervised Learning}
Supervised learning algorithms are designed to train models for predicting data classes using labeled datasets. In this domain, Yilmaz et al. \cite{DBLP:journals/corr/abs-1905-01039} proposed a method for training a Naïve Bayes classifier under Local Differential Privacy (LDP). This classifier aims to determine the most probable label for a new instance by maintaining the relationship between feature values and class labels during data perturbation. Yilmaz et al. achieved this by first transforming each user's data and label into a new value, followed by LDP perturbation. Similarly, Xue et al. \cite{8932446} also focused on training a Naïve Bayes classifier with LDP, utilizing joint distributions to compute conditional distributions. Additionally, Berrett and Butucea \cite{berrett2019classification} explored binary classification problems within the LDP framework.

High dimensionality presents a significant challenge in training classifiers with LDP, often leading to increased time costs and reduced accuracy. Traditional solutions like Principal Component Analysis (PCA) \cite{kung2014kernel} are commonly used for dimensionality reduction, but effective methods compatible with LDP in machine learning require further exploration. Another common approach in LDP-based model learning involves partitioning users, as seen in the work of Xue et al. \cite{8932446}, who divided users into groups for different calculations. However, this simple partitioning can compromise estimation accuracy. Therefore, the field of supervised learning with LDP is still evolving and requires more in-depth research.

\subsection{Empirical Risk Minimization}
Empirical risk minimization (ERM) in machine learning refers to the process of calculating an optimal model from a set of parameters by minimizing the expected loss \cite{chaudhuri2011differentially}. The loss function $\mathcal{L}(\theta ; x, y)$, parameterized by $x$ and $y$, maps the parameter vector $\theta$ to a real number.
ERM is applicable to various learning tasks like logistic regression, linear regression, and support vector machine (SVM) by selecting appropriate loss functions. Existing literature discusses both interactive and non-interactive models under Local Differential Privacy (LDP) for natural learning problems, with the interactive model offering better accuracy but at the cost of higher network delay and weaker privacy guarantees. In contrast, the non-interactive model is more robust and practical in most settings.

Smith et al. \cite{smith2017interaction} pioneered the study of interaction in LDP for natural learning problems, highlighting the network delays caused by sequential information exchange in convex optimization. They explored the necessity of interactivity for optimizing convex functions and introduced new algorithms that are either non-interactive or require minimal interaction. Building on this, Zheng et al. \cite{zheng2017collect} developed more efficient algorithms using Chebyshev expansion in a non-interactive LDP setting, achieving quasi-polynomial sample complexity.

However, the sample complexity in high dimensions remains a challenge, as noted by Wang et al. [32], who proposed LDP algorithms with error bounds based on Gaussian width. They improved upon Smith et al.'s work, but the sample complexity was still exponential with dimensionality. Their subsequent work \cite{wang2019differentially} enhanced this to quasi-polynomial complexity. For generalized linear models (GLM), Wang et al. \cite{wang2021estimating} demonstrated that LDP algorithms can attain fully polynomial sample complexity when the GLM's feature vector is sub-Gaussian with a bounded $\ell_{1}$-norm. Additionally, Wang and Xu \cite{wang2020principal} addressed the principal component analysis (PCA) problem under non-interactive LDP and established the lower bounds of the minimax risk in both low and high dimensional settings.

Classical machine learning models under LDP are typically built using ERM, solved via stochastic gradient descent (SGD). This approach is applied to linear regression, logistic regression, and SVM classification. With LDP, the SGD algorithm updates the parameter vector $\theta$ iteratively, ensuring privacy by perturbing each gradient into a noisy version before aggregation. In the context of data center networks (DCN), Fan et al. \cite{fan2020privacy} examined LDP-based support vector regression (SVR) classification for cloud-supported data centers, employing the Laplace mechanism for LDP. Yin et al. \cite{yin2019local} approached LDP-based logistic regression classification through steps of noise addition, feature selection, and logistic regression. LDP has also been applied to online convex optimization by van2019user, ensuring privacy in adaptive online learning without disclosing parameters. Furthermore, Jun and Orabona \cite{jun2019parameter} investigated the parameter-free SGD problem under LDP, proposing BANCO, which achieves the convergence rate of tuned SGD without multiple runs, thereby reducing privacy loss and conserving the privacy budget.

\subsection{Reinforcement Learning}
Reinforcement learning, a key component in artificial intelligence (AI), facilitates interactive model learning for an agent but is susceptible to attacks that can cause significant privacy breaches \cite{pan2019you}.

Addressing these concerns, Gajane et al. \cite{gajane2018corrupt} were among the first to explore the multi-armed bandits (MAB) problem in the context of Local Differential Privacy (LDP). They developed an LDP-compliant bandit algorithm targeted at arms with Bernoulli rewards. Building on this, Basu et al. \cite{basu2019differential} introduced a set of foundational privacy definitions for MAB algorithms, incorporating both the graphical and LDP models. Their work provided insights into both distribution-dependent and distribution-free regret lower bounds.

In the realm of distributed reinforcement learning, Ono and Takahashi \cite{ono2020locally} proposed the Private Gradient Collection (PGC) framework. PGC enables the private learning of a model using noisy gradients, where each local agent submits perturbed gradients in compliance with LDP to a central aggregator, which then updates the global parameters. Furthering this research, Ren et al. \cite{ren2020multi} focused on regret minimization in MAB problems under LDP, establishing a tight regret lower bound. They introduced two algorithms that achieve LDP, one based on Laplace perturbation and the other on Bernoulli response.

Reinforcement learning is a pivotal aspect of AI \cite{li2017deep}, and LDP emerges as a promising technique to protect sensitive information within this field. However, the application of LDP in reinforcement learning is still in its early stages, indicating a significant potential for future developments.

\subsection{Deep Learning}
Deep learning has significantly impacted fields like natural language processing and image classification. Yet, it is vulnerable to adversaries who can inject malicious algorithms during training to approximate and extract sensitive user information \cite{osia2020hybrid}.

To mitigate these risks, Arachchige et al. \cite{arachchige2019local} developed an LDP-compliant mechanism called LATENT for deep learning models, such as convolutional neural networks (CNNs). LATENT, like other LDP frameworks, incorporates a randomization layer (LDP layer) to protect against untrusted learning servers. A primary challenge in applying LDP to deep learning is the extremely high sensitivity, which in LATENT is expressed as $l r$, where $l$ is the binary string length and $r$ is the number of neural network layers. To overcome this, Arachchige et al. enhanced the One-time Unary Encoding (OUE) \cite{tianhao2}, which has a sensitivity of 2, by introducing Modified OUE (MOUE). MOUE offers greater control over the randomization of $1$s and increases the likelihood of retaining 0 bits in their original state. They also proposed the Utility Enhancing Randomization (UER) mechanism, which further improves the utility of randomized binary strings.

Additionally, Zhao \cite{zhao2018distributed} explored distributed deep learning under Differential Privacy (DP) using the teacher-student paradigm, allowing for personalized privacy parameter choices for each distributed data entity under LDP. Xu et al. \cite{xu2019edgesanitizer} implemented LDP in a deep inference-based edge computing framework, enabling the private construction of complex deep learning models. Despite these advancements, research in deep learning with LDP is still in its infancy. Future research is necessary to enhance privacy protection, address high dimensionality challenges, and improve accuracy.

\subsection{Federated Learning}
Federated learning (FL), a pivotal technology for advancing modern artificial intelligence (AI), offers a collaborative learning framework for multiple data owners or parties \cite{yang2019federated, li1907federated}. While FL adeptly balances utility and privacy in machine learning \cite{zheng2020preserving}, transmitting or exchanging model parameters can still lead to privacy breaches. Hence, Local Differential Privacy (LDP) has been widely implemented in FL systems to provide robust privacy assurances, as evidenced in applications like smart electric power systems \cite{zhao2020local} and wireless channels \cite{seif2020wireless}. Early studies in FL predominantly used global Differential Privacy (DP) \cite{geyer2017differentially} for safeguarding sensitive information. However, given FL's distributed nature, LDP is more suited to FL systems. Truex et al. \cite{kim2021federated} integrated LDP into FL for the joint training of deep neural networks, offering a method that efficiently handles complex models and defends against inference attacks while providing personalized LDP. Additionally, Wang et al. \cite{wang2018empirical} introduced FedLDA, an LDP-based latent Dirichlet allocation (LDA) model within an FL framework. FedLDA uses a novel random response with prior, enhancing privacy without dependency on the dictionary size and improving accuracy through adaptive, non-uniform sampling.

Bhowmick et al. \cite{bhowmick2018protection} proposed a relaxed optimal LDP mechanism for private FL to enhance model fitting and prediction. Li et al. \cite{li2019differentially} developed an efficient LDP algorithm for meta-learning, applicable to personalized FL. In federated Stochastic Gradient Descent (SGD), LDP has been employed to protect gradient privacy. However, as the dimensionality $d$ increases, the privacy budget depletes rapidly, and noise escalation leads to diminished model accuracy for large $d$. To address this, Liu et al. \cite{liu2020fedsel} introduced FedSel, which selects only the top-$k$ most important dimensions, stabilizing the learning process. Sun et al. \cite{sun2020ldp} suggested improving accuracy by reducing noise variance through splitting and shuffling, thus mitigating privacy degradation.

Naseri et al. \cite{naseri2020toward} recently presented an analytical framework to empirically evaluate the effectiveness of LDP and CDP in FL. Their findings indicate that both LDP and global DP can counter backdoor attacks but are less effective against property inference attacks.

\section{Conclusion}

In the field of data privacy, Local Differential Privacy (LDP) is an essential tool for protecting user data. Our research covers a wide range of LDP mechanisms and variations, each designed for specific challenges and data types, from BLENDER for categorical data to specialized versions like CLDP for small datasets. We also explore how LDP is applied in different machine learning areas, such as supervised learning, deep learning, and federated learning, highlighting its adaptability. The choice of LDP variant is critical, depending on the data and privacy requirements. Overall, LDP's versatility makes it a key player in ensuring data privacy across various contexts.

\bibliographystyle{IEEEtran}
\bibliography{ref}

\end{document}